\documentclass[8.5pt,twoside,twocolumn]{article}
\usepackage[super,sort&compress,comma]{natbib} 
\usepackage{mhchem}
\usepackage{times,mathptmx}
\usepackage{sectsty}
\usepackage{balance} 

\usepackage{graphicx} 
\usepackage{lastpage}
\usepackage[format=plain,justification=raggedright,singlelinecheck=false,font=small,labelfont=bf,labelsep=space]{caption} 
\usepackage{fancyhdr}
\begin{document}

\thispagestyle{plain}
\fancypagestyle{plain}{
\fancyhead[L]{\includegraphics[height=8pt]{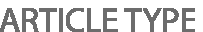}}
\fancyhead[C]{\hspace{-1cm}\includegraphics[height=20pt]{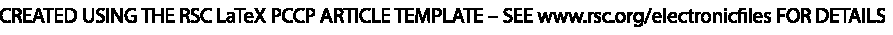}}
\fancyhead[R]{\includegraphics[height=10pt]{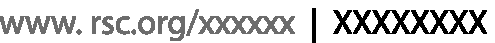}\vspace{-0.2cm}}
\renewcommand{\headrulewidth}{1pt}}
\renewcommand{\thefootnote}{\fnsymbol{footnote}}
\renewcommand\footnoterule{\vspace*{1pt}%
\hrule width 3.4in height 0.4pt \vspace*{5pt}} 
\setcounter{secnumdepth}{5}

\makeatletter 
\def\subsubsection{\@startsection{subsubsection}{3}{10pt}{-1.25ex plus -1ex minus -.1ex}{0ex plus 0ex}{\normalsize\bf}} 
\def\paragraph{\@startsection{paragraph}{4}{10pt}{-1.25ex plus -1ex minus -.1ex}{0ex plus 0ex}{\normalsize\textit}} 
\renewcommand\@biblabel[1]{#1}            
\renewcommand\@makefntext[1]%
{\noindent\makebox[0pt][r]{\@thefnmark\,}#1}
\makeatother 
\renewcommand{\figurename}{\small{Fig.}~}
\sectionfont{\large}
\subsectionfont{\normalsize}

\fancyfoot{}
\fancyfoot[LO,RE]{\vspace{-7pt}\includegraphics[height=9pt]{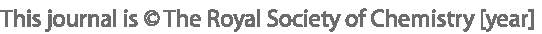}}
\fancyfoot[CO]{\vspace{-7.2pt}\hspace{12.2cm}\includegraphics{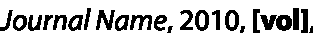}}
\fancyfoot[CE]{\vspace{-7.5pt}\hspace{-13.5cm}\includegraphics{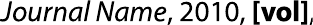}}
\fancyfoot[RO]{\footnotesize{\sffamily{1--\pageref{LastPage} ~\textbar  \hspace{2pt}\thepage}}}
\fancyfoot[LE]{\footnotesize{\sffamily{\thepage~\textbar\hspace{3.45cm} 1--\pageref{LastPage}}}}
\fancyhead{}
\renewcommand{\headrulewidth}{1pt} 
\renewcommand{\footrulewidth}{1pt}
\setlength{\arrayrulewidth}{1pt}
\setlength{\columnsep}{6.5mm}
\setlength\bibsep{1pt}

\twocolumn[
  \begin{@twocolumnfalse}
\noindent\LARGE{\textbf{Aspects of Jamming in Two-Dimensional Frictionless Systems }}
\vspace{0.6cm}

\noindent\large{\textbf{C. Reichhardt\textit{$^{a}$} and C.J. Olson Reichhardt
\textit{$^{a}$}}}\vspace{0.5cm}

\noindent\textit{\small{\textbf{Received Xth XXXXXXXXXX 20XX, Accepted Xth XXXXXXXXX 20XX\newline
First published on the web Xth XXXXXXXXXX 200X}}}

\noindent \textbf{\small{DOI: 10.1039/b000000x}}
\vspace{0.6cm}

\noindent \normalsize{
In this work we provide an overview of jamming transitions in two dimensional 
systems focusing on the limit of frictionless particle interactions in the
absence of thermal fluctuations.
We first discuss jamming in systems with short range repulsive interactions, 
where the onset of jamming
occurs at a critical packing density and where certain quantities 
show a divergence indicative of critical behavior. 
We describe how aspects of the dynamics change as the jamming density is approached 
and how these dynamics can be explored using externally driven probes.  
Different particle shapes
can produce jamming densities much lower than those observed for disk-shaped
particles, and we show how jamming exhibits fragility for some shapes while
for other shapes this is absent.
Next we describe the effects of long range interactions
and jamming behavior in 
systems such as charged colloids, vortices in type-II superconductors, and
dislocations. 
We consider the effect of adding obstacles to frictionless jamming systems
and discuss connections between this type of jamming and systems that 
exhibit depinning transitions.  
Finally, we discuss open questions such as whether the jamming transition 
in all these 
different systems can be described by the same or a small subset of 
universal behaviors, as well as future directions for studies of jamming
transitions in two dimensional systems, such as jamming in self-driven or
active matter systems.
}
\vspace{0.5cm}
 \end{@twocolumnfalse}
  ]

\section{Introduction}

\footnotetext{\textit{$^{a}$~Theoretical Division, Los Alamos National Laboratory, Los Alamos, New Mexico 87545, USA. Fax: 1 505 606 0917; Tel: 1 505 665 1134; E-mail: reichhardt@lanl.gov}}

Jamming is defined to occur when
a system consisting of a collection of interacting particles  
passes
from some type of flowing state with liquid-like properties 
to a stuck or rigid state 
with solid-like properties.  
An everyday example of jamming 
occurs in the flow of salt or coffee beans down 
a funnel. 
Even when the aperture of the funnel is significantly larger than the 
diameter of any individual flowing particle, the system can undergo a 
transition from flowing to clogged. 
At low particle density, the flow in such systems normally continues
unimpeded; however, when the density of particles is high enough, a jamming
event can occur when enough particles come into contact with each other to
form a rigid structure that halts further flow.
The jamming in this case is caused by a collective configuration of 
particles spanning the gap between the funnel walls.
Other types of jammed states can also occur in confined and 
unconfined geometries for increasing particle densities. 
The formation of these jammed states differs from
the freezing of a liquid into a crystalline solid since the jammed
states are generally spatially disordered and  
can have properties 
that are very different from those of a typical crystalline solid. 
For example, they may only be jammed against certain 
types of perturbations but not others, or they may act like a solid
in some directions and a liquid in others, so that they often have aspects
of fragility and may only be marginally rigid states.
An example of such
fragility
appears when a small tap or shake applied to a clogged salt shaker or 
to jammed particles in a hopper 
allows the system to start flowing again after a jamming event.              
Another aspect that makes jammed systems differ from typical 
crystallized solids is that jammed systems are generally 
nonthermal. 
This means
that although a system may enter a jammed state at a given density,
there is no guarantee that this configuration represents 
the lowest energy state of the system. 
Jammed systems may also appear at first sight to be similar 
to systems that form a  glass state; however, in
many of the commonly studied jamming systems such as granular media,
there is a complete absence of 
any type of thermal fluctuations, and
the question
of whether aspects of jamming can be connected directly to a glass 
remains open \cite{1}. 

The basic idea that increasing the density of a collection of 
interacting particles can cause a change from some type of flowing state
to a jammed state
applies to a remarkably wide variety of systems. Examples 
include highway traffic, where when the density of cars is high enough the 
system can become congested or jammed \cite{2,3}, 
dense flow of people through a doorway in emergency situations 
\cite{4}, and the flow of motor proteins through a cell, where at high
enough density the proteins get in each others' way and impede the flow 
\cite{5}. In materials science
it is known that the higher the dislocation density in a material becomes,
the harder it is to cause the dislocations to 
move at yielding \cite{6}.
There are also systems that exhibit depinning transitions where 
a collection of particles interact with each other as well as 
with pinning sites in a substrate.  In this case
some of the  particles can be trapped directly 
by the pinning sites, and these pinned particles can 
then impede the motion of other particles not directly 
trapped at pinning sites, causing a flowing system to become jammed  
\cite{7}. 

Although all these systems can have vastly different length 
scales and different types of interactions, 
there has been growing interest in determining whether they can be
described by the general concept of jamming, 
and whether jamming
has universal characteristics that are independent of the different 
microscopic aspects of the systems. 
Even though jamming and clogging phenomena have been familiar ever since
humans began to handle particulate 
matter, the idea that jamming could be a distinct transition that 
can be understood using  methods of statistical mechanics 
was only suggested recently.    
In 1998, Liu and Nagel proposed a scenario in which a
loose assembly of particles such as granular media, bubbles, or emulsions 
could have a unique point, termed point J, as a function of increasing 
density 
that marks the transition to a jammed state \cite{1}. They
also suggested that the physics and 
dynamics near point J could remain robust 
to perturbations such as the application of a load,
where below a critical load the 
material can be said to be jammed, while above yield the
material is unjammed \cite{1,8}. 
As a function of temperature, the high temperature state 
of system will form a liquid that flows or is unjammed, 
while at lower temperatures the system freezes 
into a jammed solid-like phase \cite{1}. 
Around the same time, Cates {\it et al.} also proposed that certain 
systems that are 
solid or in a jammed state
represent a new class of materials called fragile materials 
that only behave like a solid under certain perturbations
and that have a strong memory dependence \cite{9}. 
Since these simple but elegant ideas were proposed, 
there has been a tremendous growth in jamming research,
and the idea of jamming as a 
distinct type of transition 
is now being applied to an ever wider class of materials, 
including quantum systems \cite{10}. 

Here we focus on aspects of jamming in two dimensions (2D), primarily in
systems where there is an absence of friction and 
where thermal effects are negligible. 
Although this is only a small subset of the systems exhibiting
jamming that are currently being 
studied,
we show that even these simple 2D systems 
exhibit a remarkable variety of behaviors, 
and that there are many open problems and 
new directions to explore in this area.  

\section{Nature of the Transitions and Correlation Lengths}

Ever since a scenario for a unified picture of jamming transitions 
was first proposed in Ref.~\cite{1}, one of the most 
popular theoretical and numerical models for investigating jamming 
in 2D systems has been frictionless disks interacting
with harmonic or Hertzian short range repulsion with a well defined range, 
where the disks have two different radii $r_A$ and $r_B$
with $r_{A}/r_{B} = 1.4$ \cite{8}. A 2D assembly of monodisperse
disks forms a triangular lattice at a density close to $0.9$, 
and for this reason the bidisperse disk systems are used 
since they avoid crystallization.  

A key question is whether point J, the transition into a jammed state,
has properties 
similar to those found for
phase transitions observed in equilibrium and certain nonequilibrium systems. 
The first complication is that 
phase transitions are typically identified though the 
breaking of some form of spatial symmetry; however, 
jamming systems are typically disordered or amorphous whether jammed
or not.
Since there is little or no change in the
spatial structure of the system across jamming, 
identifying the proper order parameter can be challenging. 
For second order or continuous phase transitions, there is
some length scale that shows a divergence as the critical point 
between the two phases is approached.  
In the Ising model, the two phases are a ferromagnetic ordered state and a 
paramagnetic or disordered state, 
and at the transition between these phases, various quantities 
become scale free \cite{11}. 

If the system exhibits a continuous phase transition, the correlation
length $\xi$ grows as a power law as the density $\phi$ approaches
the jamming density $\phi_j$,
\begin{equation}
\xi \propto (\phi_{j} - \phi)^{\nu}. 
\end{equation}
In principle, if the scaling behavior can be found and 
the critical exponents such as $\nu$ can be identified accurately, 
it would be possible to
determine the exact nature of the transition and see if it 
falls into one of the already known universality classes such as 
Ising model, directed percolation, or Potts \cite{11}.
Once the nature of the transition has been identified in 
one system that exhibits jamming, 
other systems with different microscopic details that also exhibit jamming
can be examined
to see if they fall into the same universality class or to understand
why they do not.

Accurately measuring critical exponents
can be difficult even in equilibrium systems, 
where a variety of complications and subtleties can arise. 
Such complications
include understanding how close it is necessary to be to 
the critical point to obtain accurate scaling 
and how large the system must be to
avoid finite size effects \cite{12}. 
The jamming systems have the extra complication of generally being
out of equilibrium; 
however, even nonequilibrium systems can exhibit continuous phase 
transitions \cite{13}. 
Despite the significant number
of theoretical and numerical studies on nonequilibrium 
systems that exhibit phase transitions \cite{13}, only
in recent years have experimental systems been realized 
that exhibit clear nonequilibrium phase transitions \cite{14.1,28}. 	   

An issue that often arises in systems that exhibit jamming is 
that the jamming density obtained for one type of
preparation method can differ from that obtained by preparing the system
in a different manner \cite{14,15}.
Furthermore, the measurable quantities in the system can differ depending
upon whether
the jamming transition is approached
from below by increasing the density $\phi$ or from above by starting
in an already jammed high density state 
and decreasing $\phi$ to the jamming density $\phi_{j}$. 
For example, below $\phi_j$ it is possible to arrange the
disks such that no two disks are in contact, making it impossible to
measure vibrational spectra, whereas in the jammed state above
$\phi_j$ where the system acts as a solid, vibrational spectra can
be analyzed.
Multiple length scales may come into play, but these scales might be
distinguishable only in very large systems \cite{16}.
It is also possible that
jamming
exhibits
a mixture of first and second order 
features or that it is weakly first order \cite{16.N}.   

The initial studies of jamming focused on finding 
evidence that jamming exhibits some critical properties 
associated with continuous phase transitions, and
concentrated on identifying quantities that exhibit scaling  
near a critical density or a critical load.
O'Hern {\it et al.} showed that in a frictionless bidisperse disk system
several quantities scale upon approaching a critical jamming
density of $\phi_j\approx 0.844$ \cite{8}, and suggested
that one possible way to directly examine a diverging length scale 
as jamming is approached would be 
to push a single probe particle through a random assembly of disks
and measure the perturbation distance in the surrounding disks. 
Well below the jamming density, this perturbation length 
scale would be small, but the
perturbation length would diverge as the system approaches 
the critical jamming density at which it becomes marginally rigid. 
Drocco {\it et al.} \cite{17} subsequently simulated 
dragging a probe particle
through a 2D bidisperse frictionless disk assembly, 
where overdamped motion of the particles was assumed.
Figure~1, adapted from Ref.~\cite{17}, 
shows the location of the probe particle, 
the unperturbed particles, and
the particles that are experiencing a force due to the motion of
the probe particle.  
In Fig.~1(a), at $\phi=0.656$, well below $\phi_j$, 
an empty trail forms behind the probe particle.
The trail does not refill with particles because there is no 
pressure in the unjammed phase.  
As the probe particle moves through the system, 
it pushes other particles out of the way, and these particles
push other particles they encounter, and so forth, producing a perturbed area
that varies in size depending on the density of the packing.
In Fig.~1(b) at $\phi = 0.8$,
there is still a void behind the probe particle since the sample is
unjammed; 
however, the extent of the temporarily perturbed region has increased.
In 
Fig.~1(c) at $\phi = 0.83$, the system is just at the 
jamming density for this size of simulation area, and the motion 
of the probe particle immediately propagates throughout the sample,
indicating that the probe particle is dragging all of the other particles
along with it.
Another signature that the system has jammed is the disappearance of the
trailing void region behind the probe particle. 
Experiments in 2D disk assemblies with a driven probe particle 
also found that 
at low density a void or cavity opens behind the probe particle, while 
near the jamming density the void disappears \cite{18}. 

\begin{figure}
\centering
\includegraphics[width=0.5\textwidth]{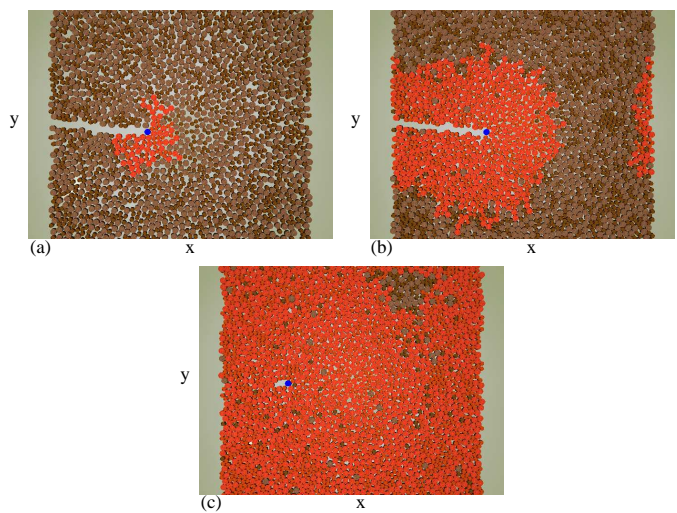}
\caption{Snapshots of a jamming length scale probe showing the perturbation
caused by a single driven particle.
Blue: Probe particle driven with a constant 
force in the positive $x$-direction;  brown: bidisperse background particles;
red: the cluster of background particles in force contact with the driven
particle at this instant in time.
(a) Unjammed state at $\phi = 0.656$. 
(b) Unjammed state at $\phi = 0.8$, where the perturbation has
crossed the periodic boundary conditions in the $x$ direction. 
In (a) and (b) the dragged particle leaves behind an empty trail. 
(c) Jammed state at $\phi = 0.83$, where the dragged particle no longer
leaves a trail and the perturbation wraps around the periodic
boundaries.
The figure is adapted from ref.~\cite{17}.
}
\label{fig:1}
\end{figure}

\begin{figure}
\includegraphics[width=3.5in]{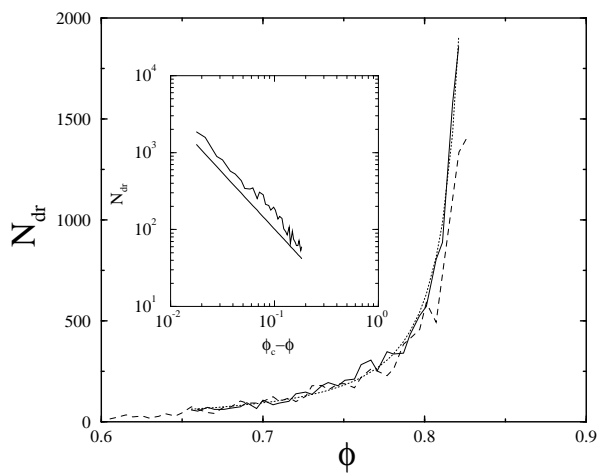}
\caption{$N_{dr}$, the average number of grains 
dragged by the probe particle, vs $\phi$, showing a divergence near
$\phi_{j}\approx 0.84$.  The inset shows that this divergence 
has a power law form suggestive of critical behavior.
The figure is adapted from Ref.~\cite{17}.
} 
\label{fig:2}
\end{figure}

Let the velocity of a probe particle in an overdamped 
medium under a constant force $F_{D}$ in the absence of any other particles
be $V_{0}$.
When placed inside a packing above the jamming density,
the average probe particle velocity $\langle V\rangle$ 
drops to $V_0/N$, where $N$ is the total
number of grains in the system.  In the thermodynamic limit, 
$\langle V\rangle \rightarrow 0$.
At a particle density just below jamming, 
$\langle V\rangle=V_{0}/N_{dr}$,
where $N_{dr}<N$ is the number of grains dragged with the probe particle. 
In Fig.~2, adapted from Ref.~\cite{17}, we plot $N_{dr}$ 
as a function of $\phi$, showing a divergence at  
$\phi_{j} \approx 0.84$.  
The inset of Fig.~2 shows that the divergence has 
a power law form $N_{dr} \propto (\phi_{j} -\phi)^{\delta}$.
Assuming that the cluster of dragged particles is circular in shape,
we have
$\delta = \nu(d +2)$ with $d = 2$, and using the value of $\delta$
extracted from the figure, we obtain
the critical exponent $\nu =0.71$ \cite{17}.  

By examining different quantities,
O'Hern {\it et al.} 
found the value $\nu = 0.6-0.7$ \cite{8}. 
A number of other numerical studies have been performed on the same
bidisperse disk system in order to extract critical exponents.
Olsson and Teitel conducted shear simulations and found
$\nu = 0.6 \pm 0.1$ \cite{19}, Heussinger and Barrat obtained
$\nu = 0.8-1.0$ in quasistatic shear simulations \cite{20}, 
and relaxation studies performed by
Head gave $\nu \approx 0.6$ \cite{21}.  More recently,
Vagberg {\it et al.} carried out extensive finite size size 
scaling on very large 2D systems,
and argue that corrections to scaling are 
very important, producing more accurate
calculations of  $\nu \approx 1.0$ \cite{12}. 
There are also a number of theoretical models that give diverging correlation
lengths as point J is approached from the high density side.
Wyart {\it et al.} obtain $\nu=1/2$
using a cutting argument\cite{22}, 
while Silbert {\it et al.} argue for two different exponents,
$\nu_{L} = 0.48$ for the longitudinal length and
$\nu_{T} = 0.24$ for the transverse length \cite{16}.
Using k-core percolation, Schwarz {\it et al.} 
found two exponents of $1/2$ and $1/4$ \cite{23}, 
while field theoretical
studies of Henkes and Chakraborty give $\nu = 1/4$ \cite{24}. 
Recently Goodrich {\it et al.} \cite{25} presented results 
suggesting that the cutting length found by Wyart 
{\it et al.} \cite{22} can be understood as a rigidity length and that it
is correlated with the longitudinal length 
found by Silbert {\it et al.} \cite{16}.     
Waitukaitis {\it et al.} \cite{wait} 
performed an experimental study of a dynamical
jamming front induced by uniaxial compression of a packing with a rake.
Ahead of the wake is a densification front that has a width which diverges as
the jamming density is approached.  Analysis of the front width gives
an exponent $\nu=0.65$.

Other studies also find scaling in 
other quantities \cite{24.N,24.N1} such as
the contact number $Z$, where 
$Z - Z_{j} = Z_{0}(\phi - \phi_{j})^\eta$, 
with  $\eta = 0.5$ in 2D \cite{8,ohern2002,katgert2010}. 
This value has also been observed in experiment \cite{26}.
One interesting feature observed by O'Hern 
{\it et al.} \cite{8} is that  many of the scaling exponents
for the different quantities change depending upon the detailed nature
of the interparticle interactions.   This is different from the behavior
typically observed at critical points, where
systems that are in the same universality class 
have the same exponents regardless of the microscopic interactions
between the particles.     
Diverging quantities near jamming were also observed in 2D
lattice models \cite{newhall2013}.

Recently, a new study described jamming in sheared 2D systems 
in terms of a transition from chaotic behavior
below jamming to non-chaotic behavior above 
jamming \cite{27}. 
Here, the system exhibits chaotic 
rearrangements below jamming, but above jamming
the system becomes rigid and can no longer move in a chaotic fashion.
The authors measure
a dynamical length scale $\xi_{d}$, 
and showed that as the jamming density of $\phi_c=0.841$ is approached,
$\xi_d$ decreases as a power law,
$\xi_{d} = (\phi_c -\phi)^{-\alpha}$, with
$1.7 < \alpha < 2.1$. 
This exponent is considerably different than 
the other diverging length scale exponents 
discussed previously; however, 
the 
length $\xi_d$ may be relevant to the dynamical rather than
the static properties of the jamming transition, and may not be closely
related to the 
diverging length scales considered in the other studies.
The association of jamming with a transition from chaotic to non-chaotic
motion 
provides 
another definition for the onset of jamming. 
It would interesting to apply a similar 
approach to jamming systems in which friction between the particles 
is relevant and where the jamming density is
considerably lower than $\phi=0.841$. 
Other studies in 2D systems also find growing
dynamical length scales as jamming is approached \cite{keys2007,heu2010}.

The idea of understanding jamming in terms of dynamical processes 
can also be studied by applying a periodic drive 
to the sample and observing whether the system reaches a reversible
state. 
This approach was used by Corte {\it et al.} 
for colloids well below jamming, where a transition from reversible to
irreversible behavior was linked to a nonequilibrium phase transition
\cite{28}.           
Schreck {\it et al.} \cite{29} recently studied 
a frictionless disk system using the 
periodic shear protocol of Ref.~\cite{28} 
and found several different dynamical regimes below $\phi_{j}$, 
including
a point reversible regime, transient reversible regime, 
and an irreversible regime. 
This
work suggests that there may be additional length scales or transitions 
at densities below $\phi_{j}$ that
come into play under driving.   
For example, Shen {\it et al.} \cite{30} found 
a percolation transition at a density $\phi_{p} < \phi_{j}$. 
The percolation 
occurs through the formation of a non-rigid system-spanning cluster 
that causes
the onset of non-trivial mechanical responses 
related to the emergence of correlated particle motions.   
If there are additional dynamical transitions below $\phi_j$, this could
be of importance in connecting jamming with glass transitions.
Olsson and Teitel found
in 2D systems that the onset of glassy behavior occurs at 
a density $\phi_{g} < \phi_j$ \cite{31}.
It would be interesting to understand whether
$\phi_{g}$ is related to the percolation density 
$\phi_{p}$ found by Shen {\it et al.} \cite{30}. 
For densities greater than $\phi_{j}$, 
studies have suggested that there could be new types of phenomena 
with different mechanical properties
in what is called the deep jamming regime, 
where certain quantities scale differently than they do 
close to point J in the marginally jammed
regime \cite{32}.
If very dense jammed systems exhibit distinct phenomena, this could
be relevant for understanding many types of jammed soft matter systems
such as emulsions \cite{33} or foams \cite{34}, where 
densities well above $\phi_j$ can be accessed.

Although there has been mounting evidence that 
point J exhibits critical properties, 
the strong variations in the values of the extracted critical exponents
may mean that more than one length scale diverges at point J, 
or that point J has properties that are different from those normally
associated with critical points.
Even for the heavily studied bidisperse disk system,
the exact nature of the jamming transition 
remains an open issue. 
Once this has been settled, the next question will be to 
determine whether the same type of transition 
exists in other jamming systems and frictional systems.  

Comparisons between theoretical predictions and jamming experiments 
can be complicated by the existence of frictional effects in the experimental
system, both in the interactions between grains and in the motion of the
grains against the containing vessel.
Typically, 2D jamming densities found in experiments \cite{18,26,35} are lower 
than the value $\phi_j=0.844$ 
obtained from simulations, and this difference
is likely due to friction. 
There are several 
possible jamming scenarios for the frictional systems. 
For example, there could be
two separate jamming transitions, a frictional jamming at lower
density as well as
a point J frictionless jamming at higher density that  could
be accessed by further compressing the system.
There are some studies 
indicating that frictional jamming 
can have the same nature as frictionless jamming \cite{36}.

\section{Fluctuations Near Jamming}   

Another approach to understanding jamming is  
by analyzing changes in dynamically fluctuating quantities.  
If the fluctuations increase near jamming, this would be consistent
with 
the idea that jamming has the properties of a second order transition,
with scale free fluctuations of all sizes at the critical point. 
Geng and Behringer studied the fluctuating drag force on a probe particle
dragged at constant velocity through packings of different densities \cite{37}.
They found a characteristic packing density
$\phi_j \approx 0.653$ below which
the force fluctuations become very small. 
This characteristic packing density   
is considerably below the ideal point J density $\phi=0.844$,
indicating that friction plays an important role in their system.
Nonetheless, 
as the packing density increases toward $\phi_j$ in 
this study, 
the average force 
$\langle F\rangle$ on the probe increases
according to a power law, $\langle F\rangle \propto (\phi_j -\phi)^{1.53}$,
while 
the distribution of force fluctuations develops
increasingly long tails 
that can be fit to an exponential distribution. 
The power spectrum of the force fluctuation time series 
has a $1/f^2$ 
form and the overall noise power increases
with increasing density. 
The distribution of the size of avalanche-like jumps in
force 
is exponential \cite{37}.    

\begin{figure}
\includegraphics[width=3.5in]{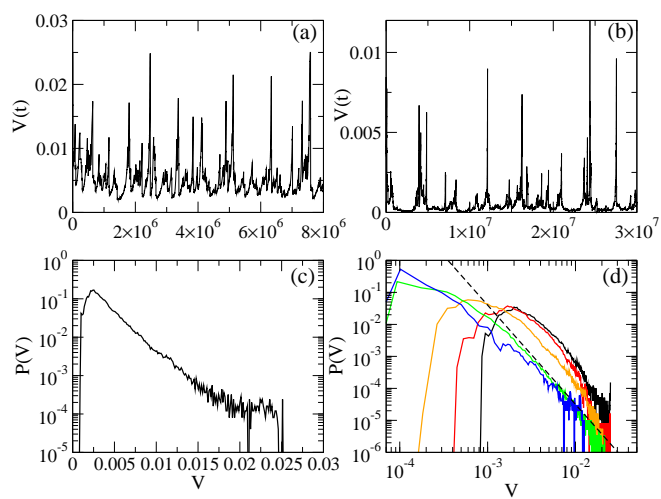}
\caption{ (a,b) The time series of the velocity of the probe particle 
$V(t)$ from simulations
of a 2D bidisperse disk system at
(a) $\phi = 0.71$ and (b) $\phi = 0.8414$.
(c,d) The velocity distribution functions 
$P(V)$ for a probe particle driven through a bidisperse disk system.
(c) At $\phi = 0.807$, $P(V)$ is exponential. 
(d) $\phi=0.823$ (black), 0.833 (red), 0.8395 (orange), 0.8414 (green),
and 0.8427 (blue).  At densities near $\phi_j=0.844$, there is
a crossover to a power law velocity distribution.
The dashed line indicates a power law fit to the $\phi=0.8427$ curve with
$\alpha = -2.75$.
The figure is adapted from Ref.~\cite{38}.
}
\label{fig:3}
\end{figure}

In simulations of a probe particle driven with constant applied force
through a frictionless packing,
the velocity
fluctuations also become increasingly intermittent upon approaching
$\phi_j = 0.844$ \cite{17,38}, as shown in Fig.~3(a,b) at $\phi = 0.71$ and 
$\phi = 0.8414$ \cite{38}. 
The average velocity  $\langle V\rangle$ of the probe
particle decreases linearly with increasing $\phi$ 
until $\phi \approx 0.83$, above which 
$\langle V\rangle$ drops off more rapidly 
before reaching zero for $ \phi > 0.844$ \cite{38}.
The distribution of the velocity fluctuations $P(V)$
for $\phi < \phi_{j}$ has an exponential tail, 
illustrated in Fig.~3(c) for $\phi=0.807$ \cite{38}, similar 
to that found experimentally in Ref.~\cite{37}. 
As the density increases, there is a crossover in the distribution to a
power law form, as illustrated in Fig.~3(d), where the
$\phi = 0.8414$ curve is fit by a power law with exponent
$\alpha = -2.75$ \cite{38}.  
It is possible that power law distributions were not observed
experimentally since they appear only in the critical region
very close to the frictionless point J.
Measurements of the power spectra of the probe particle velocity fluctuations
in experiments and simulations
show a characteristic knee shape for
$\phi < \phi_{j}$ 
and $1/f^2$ behavior at high frequency \cite{37,38}.
Very close to $\phi_{j}$, the form of the spectrum changes to 
a low frequency  $1/f^{1.1}$ characteristic with a persistent
$1/f^2$ shape at high frequency. 
The appearance of $1/f$ noise 
is consistent with the occurrence of
large scale rearrangements on length scales up to the entire system size,
which produce
excess low frequency noise. 
For packings with $\phi > \phi_{j}$, which can be simulated using 
harmonic particle-particle interactions,
a finite threshold force $F_{c}$ must be applied in order to 
induce local plastic rearrangements of the jammed grains and move the
probe particle through the packing, and the value of
$F_{c}$ increases  with increasing $\phi$ \cite{38}.
At  $\phi = 0.844$,  the velocity versus vs force curves   
scale as  
$\langle V\rangle \propto (F- F_{c})^{\beta}$ 
with $\beta \approx 1.5$, while for $\phi > \phi_{j}$ the exponent changes
to $\beta = 0.5-1.0$, 
indicating a modification in the dynamics of the probe particle. 
Studies of the plastic depinning of interacting particles driven over 
random substrates produce velocity-force exponents of $\beta=1.5 - 2.0$ when
the flow is strongly intermittent \cite{39},
while a single particle driven over a random or periodic substrate 
has $\beta=0.5$ \cite{40}.
This suggests that in the critical region near but below point J,
global plastic rearrangements occur as large portions of the particles
move in response to the probe particle, while above point J,
the system remains jammed and only a smaller localized region 
of particles near the probe particle undergoes plastic rearrangements when 
the probe particle depins.    

The dynamic force fluctuations near jamming have also been 
investigated experimentally using an intruder
particle driven at constant velocity \cite{41}.
A small vibration was used
to reduce the effect of friction in these experiments, 
allowing access to densities close to $\phi=0.841$.
Large scale rotational motions can occur around the intruder particle.
As the jamming density is approached in this work,
the force needed to keep the intruder particle
moving at a constant average velocity diverges, 
and the intruder particle moves in bursts of activity that become more
intermittent as the density increases.
By analyzing the distribution of the bursts and using scaling, the
authors provide strong evidence for the divergence of several quantities
and obtain a critical exponent of $\nu = 1.0$.
Event-driven simulations of frictional 2D systems 
with a driven intruder particle show that 
as the density increases, the intruder mobility decreases,
dropping close to zero for $\phi \approx 0.8$ \cite{42}. 
Kolb {\it et al.} \cite{18} conducted experiments of driven intruder particles 
and find a divergence in the mean force fluctuations $\Delta_F$
as the jamming density 
is approached, with the form $\Delta_F \propto f/(\phi_{j} - \phi)^{-1}$. 
They also find large scale circular or rotational motions
of the surrounding media near the probe particle. 

\section{Different Types of Particle Geometries}

Beyond circular disks, jamming in 2D can also be explored 
for more complex particle shapes \cite{43}, 
permitting the realization of different jamming densities and different
average particle contact numbers.
One example of such a system is ``granular polymers'' 
composed of grains strung into a finitely flexible chain with a fixed
minimum bending angle between any two grains along the chain.
Such chains have been experimentally studied with vibration
plate 
\cite{44,prentis2002} and shear \cite{46} apparatus.   
In experiments on 3D granular polymer arrangements, 
examination of the density of the system 
revealed that the final packing density decreases with
increasing chain length \cite{45}. 

\begin{figure}
\includegraphics[width=3.5in]{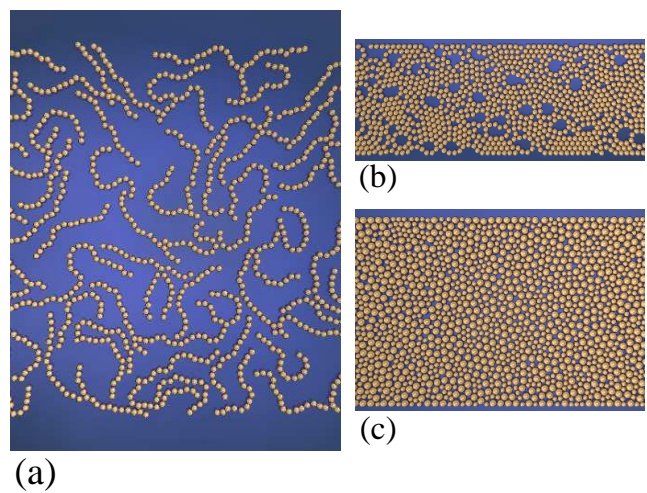}
\caption{
(a) Configuration of frictionless granular polymers below jamming for 
a system with chain length of 16.
(b) The same system compressed to the jamming density showing 
the formation of numerous voids due to the finite chain bending radius.
(c) A bidisperse disk packing at the jamming density of $\phi_j=0.84$ does
not contain any large voids.
The jamming density for the granular polymers is considerably lower than 
that of the bidisperse disks. Figure adapted from Ref.~\cite{48}.
}
\label{fig:5}
\end{figure}

Granular polymers represent an ideal system
in which the jamming density $\phi_{j}$ can be tuned  
by changing the length of the chains, and they can be used to study whether
the same types of jamming or critical behaviors found for the bidisperse disk
system still occur.
In 2D simulations of non-thermal frictionless granular polymers 
with a finite bending angle \cite{47,48},
jamming was detected by compressing two walls 
and identifying the onset of a finite pressure as well as by
conducting shear measurements. 
In Fig.~4(a) we show an image from Ref.~\cite{48} of
the unjammed state of a 2D granular polymer chain 
simulation with 16 monomers per chain,  
and in Fig.~4(b) we show the same system in a jammed configuration. 
For comparison, in Fig.~4(c) we illustrate the jammed configuration 
obtained from compression
of a bidisperse disk system.  
The jammed state of the granular polymers contains a
large number of voids that are absent in  
the jammed bidisperse disk system, 
indicating that the jamming density is 
significantly reduced in the granular polymer system.
In Fig.~5(a), adapted from Ref.~\cite{48}, we show that the pressure $P$ in
a bidisperse disk system 
increases linearly with $\phi$ for $\phi >  \phi_{j}$, as previously 
observed \cite{26}.
For granular polymers of length 6, 8, 10, 16, and 24 grains,
Fig.~5(b) shows that the jamming density drops with increasing chain length,
and that 
above jamming the linear scaling of $P$ with $\phi$ is 
replaced by an exponential scaling \cite{48}. 
In bidisperse disk systems, the contact number $Z$ scales as jamming is
approached from above with an exponent $\beta=0.5$ \cite{26},
while in the granular polymer system, $\beta=0.6 - 0.8$, and $\beta$
increases with increasing chain length \cite{48}.

\begin{figure}
\includegraphics[width=3.5in]{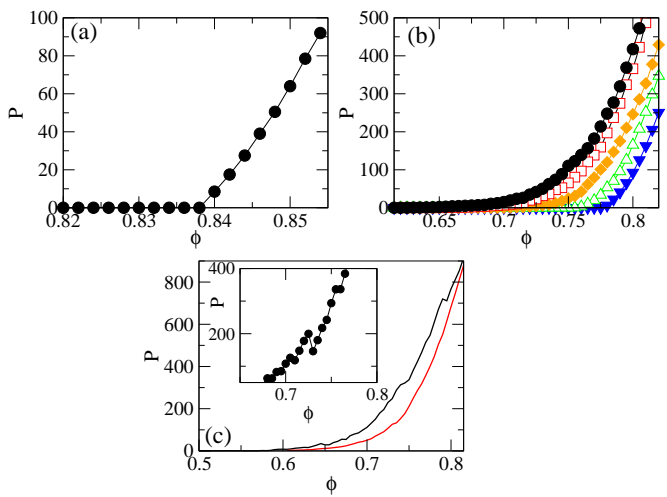}
\caption
{ (a) Pressure $P$ versus $\phi$ for the bidisperse disk system. 
Jamming occurs close to $\phi = 0.84$, and above jamming,
$P$ increases linearly with increasing $\phi$. 
(b) $P$ vs $\phi$ for granular polymer systems with chain lengths
of 6 (filled blue triangles) 8 (open green triangles), 
10 (filled orange diamonds), 
16 (open red squares), 
and 24 (filled black circles).  Above jamming, $P$ does not increase
linearly but is better fit by an exponential.
(c) $P$ vs $\phi$ for a granular polymer system with
chains of length 38 for the initial compression (red) and 
a subsequent wall compression cycle (black), 
showing strong hysteresis. 
Inset: a zoom of the main panel
showing that during the first compression cycle, sudden drops in the 
pressure occur due to plastic rearrangements in the system. 
Figure adapted from Ref.~\cite{48}.
}
\label{fig:6}
\end{figure}

There have not yet been any direct measures of a diverging length scale
in granular polymer systems, 
so the question of whether jamming for the granular polymers has a
different nature from jamming in bidisperse disk packings remains open.
Although the pressure and coordination number scale differently 
in the granular polymer and bidisperse disk systems,
this fact alone is not enough to establish that jamming in the two
systems is different and that there
is thus more than one type of jamming transition.
It is possible that extraction of a true critical length scale in the
granular polymer system would require performing measurements on
samples of extremely large size.

If bidisperse frictionless disk systems are subjected to cycles of
compression and pass through the jamming density several times,
above jamming no plastic rearrangement events occur in  
in the bulk, so there is no hysteresis in the pressure or 
coordination number.  
On the other hand, in compressive cycling of the granular polymer systems,
strong plastic rearrangements occur in the bulk of the sample during the
first cycle, producing a hysteretic response.
The plastic events occur in abrupt avalanches, as illustrated
in the inset of
Fig.~5(c) 
where a sudden pressure drop event occurs. 
Such drops are associated with the collapse of the void structures
shown in Fig.~4(b).  
Under repeated compression, the density of the granular polymers gradually
increases until eventually the system saturates to a final jamming
density $\phi_{j^\prime}$ that is significantly higher than the
$\phi_j$ measured during the first compression cycle.
At $\phi_{j^\prime}$, the granular polymer system
exhibits properties that are more similar to those found at point J 
for the bidisperse disk system; however, some voids 
always remain present in the granular polymer packing.    
The behavior of the granular polymers may be relevant to 
the case of frictional bidisperse disks, which also
jam at a much lower density than frictionless disks.
If the frictional packing is subjected to further compression,
the frictional contacts can be broken, 
producing plastic rearrangements accompanied by hysteresis \cite{zhang2010}.
The hysteresis vanishes when the system 
undergoes only elastic displacements, which could
correspond to reaching the frictional point J. 
This scenario would imply that there are two jamming transitions, 
one at lower density associated with
more fragile structures such as loops or frictional contacts, and the 
other at higher density corresponding to the frictionless or fully elastic
jamming at point J.

Beyond granular polymers, other types of 2D particle geometries can be
considered, such as dumbbell shapes.
At first glance, it might seem as if a dumbbell packing would have a
lower jamming density than the bidisperse disk system, but this is not
necessarily the case;
frictionless dumbbells with only a pointlike contact between the two dumbbell
halves
actually jam at a higher density of 
$\phi = 0.9$, and the jammed state is an ordered triangular lattice \cite{49,50}.
If the elongation of the dumbbells is varied by increasing the amount of
overlap between the two halves of each dumbbell,
the jamming density can be modified and the jammed state becomes
disordered \cite{51}.
Shreck {\it et al.} \cite{51} 
report evidence that different types of scaling occur for
a system of distorted dumbbells 
than for bidisperse disk systems. 
Mailman 
considered frictionless elliptical-shaped particles
and also found 
scaling properties different from those of bidisperse disk packings,
supporting 
the idea that such systems exhibit a different 
type of jamming transition \cite{52}. 
It would be interesting to extract the critical exponent for a 
diverging length scale in these systems to see if it is the same as
in the bidisperse disk system or if 
the elongated grain shapes can exhibit
multiple jamming behaviors. 
In addition, an analysis of the
force fluctuations in these systems in the jamming regions could show
whether
the fluctuations are exponentially distributed, like the
avalanche and velocity fluctuations of the frictional bidisperse disk
systems, or whether they are more consistent with
critical phenomena.    

\begin{figure}
\centering{
\includegraphics[width=3.5in]{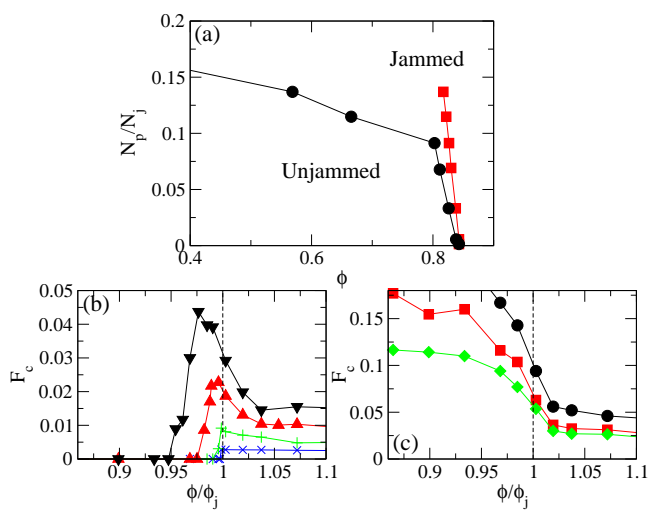}}
\caption
{ 
Jamming in a system of bidisperse disks driven through an obstacle array.
(a) Jamming diagram for $N_p/N_j$ vs $\phi$.
$N_p$ is the number of obstacles and $N_j$ is the number of disks in an
obstacle-free system at the jamming density.
$\phi$ is the disk density.
The system is defined to be jammed when the disks do not 
continuously move under an arbitrarily small
applied drift force.
Jammed states exist above and to the right of the black line, and unjammed
states are found below the black line.
The red squares indicate the value of $\phi$ above which the particles
flow as a solid. Near point J, the jamming density decreases linearly with
increasing $N_p/N_j$, while for $\phi<0.8$, the line separating the
jammed and unjammed states is much flatter. 
(b) The critical force $F_c$ required to depin the disks vs $\phi/\phi_j$,
where $\phi_j$ is the jamming density in the absence of pinning sites,
for $N_p/N_j=$ 0.00138 (blue crosses), 0.00692 (green plus signs),
0.0346 (red triangles), and 0.09267 (black triangles).
(c) $F_c$ vs $\phi/\phi_j$ for
$N_p/N_j=0.828$ (black circles), 0.415 (red squares), and 0.277 
(green diamonds).
Figure adapted from Ref.~\cite{53}.
}
\label{fig:7}
\end{figure}

\section{2D Jamming With Quenched Disorder}
There are many examples of systems where a 
collection of particles interact not only with each other but
also with 
obstacles or pinning sites. 
Such systems include 
vortices in type-II superconductors, classical Wigner crystals, 
colloids in disordered media,
and topological defects such as skyrmions \cite{39}. 
Another example is granular or colloidal media 
flowing through an array of posts, where it could be expected that as
the number of obstacles $N_p$ increases, the
density at which jamming occurs should decrease. 
By adding obstacles and a drift force
to a 2D bidisperse disk system which has a
clean jamming density of $\phi = 0.84$, it is possible to measure any
change in the jamming density and determine whether the properties of
the jamming are altered by the obstacles.
When $\phi > \phi_{j}$, a single obstacle ($N_p=1$) suffices to
pin the entire system, while 
for $\phi <  \phi_{j}$ the particles should be able to flow past a 
single obstacle. 
Now suppose $\phi$ is held fixed at a value $\phi<\phi_j$ while $N_p$,
the number of obstacles, increases.  Once $N_p$ is high
enough, the particles cease to flow and the system enters a jammed state.
In Fig.~6(a), adapted 
from Ref.~\cite{53},
we mark where the jammed and unjammed phases appear on a plot of
$N_{p}/N_{j}$ vs $\phi$, 
where $N_{j}$ is the number of disks at point J in the absence
of obstacles. 
The obstacles are modeled as pinning sites that can each capture at most one 
disk. 
A disk trapped at a pinning site can escape from the pinning site if the net
force on the disk from other disks and the drift force exceeds the pinning
force $F_p$.
In Fig.~7 the black line indicates the location in the $N_p/N_j - \phi$ plane
above which there is a nonzero critical drift force threshold $F_c$ that
must be applied in order to get the disks to move or unjam.
For values of $\phi$ above the red line, the system moves as a rigid 
elastic solid.  Between the black and red lines, the system flows plastically
when a drift force larger than $F_c$ is applied, with a coexistence of
moving and jammed regions in the sample.
As the number of obstacles increases, the value of $\phi$ at which
jamming occurs 
initially decreases linearly with $N_p$, 
until for $N_{p}/N_{j}  \geq 0.1$,
it begins to flatten off as
the system becomes much easier to pin. 
Near point J, the correlation length grows as
$\xi \propto (\phi_{j} - \phi)^{-\nu}$, 
and the system should jam when $\xi$ becomes larger than the average 
distance between pinning sites, 
$l_{p} \propto \rho_{p}^{-1/2}$, where  
$\rho_{p}$ is the pinning density. 
Thus, the line demarking the jammed state should follow
$\rho_{p} \propto (\phi_{j} - \phi)^{2\nu}$. 
The linear behavior of this line shown in Fig.~6(a)
then gives $2\nu = 1$ or $\nu = 1/2$. This exponent is 
in agreement with certain predictions \cite{16,22,23,21} 
but differs from others \cite{17,20}. 

\begin{figure}
\includegraphics[width=3.5in]{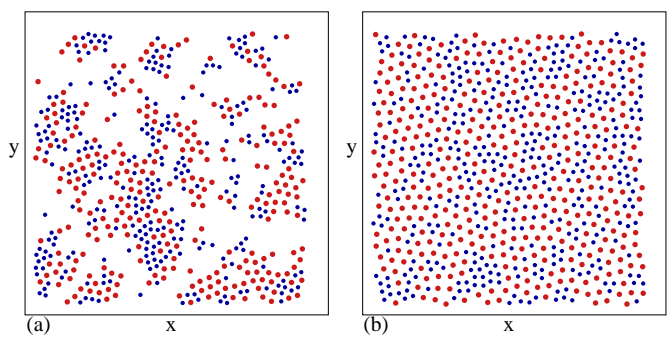}
\caption
{(a) The positions of bidisperse disks driven with a drift force through
an obstacle array. (a)
The jammed state  
at $\phi/\phi_{j} = 0.761$ and $N_p/N_j=0.415$,
where $\phi_{j}$ is the pin free jamming density. 
The disk density is highly spatially heterogeneous, with some empty
areas and other local clusters that have a density close to 
$0.844$. 
(b) The jammed state for 
$\phi/\phi_{j} = 1.04$ and $N_p/N_j=0.415$ is 
homogeneous. 
Figure adapted from Ref.~\cite{53}.
}
\label{fig:8}
\end{figure}

The jammed or pinned states that form for 
$N_{p}/N_{j} > 0.1$ and $\phi  < 0.8$ are strongly spatially heterogeneous,
as shown in Fig.~7(a) 
for a sample with $\phi/\phi_{j} = 0.761$ 
and $N_{p}/N_{j} = 0.415$ \cite{53}. 
In contrast, at the same pinning density but for
$\phi/\phi_{j} = 1.04$, Fig.~7(b) shows that the jammed state is
homogeneous. 
The heterogeneous states that form at lower densities differ from the
states near point J since they are very fragile, meaning that they are
sensitive to the direction of the drift force under which they were
prepared.
If a system in this fragile regime reaches a jammed state due to a drift
force applied along the x direction, but is then exposed to a different
drift force applied along a new direction (such as the y direction), the
disks may begin to flow again and may or may not organize into a new
jammed or pinned state.
For densities above the red line in Fig.~6(a),
the jammed state is robust under all directions of drift force and
local plastic deformations do not occur even when the drift force
exceeds $F_c$ and all of the particles begin to move.
The lower density heterogeneous states are 
better described as clogged states due to their directional fragility. 
These clogged states retain a connection
to point J since the local density of the 
clusters of disks is close to the obstacle-free jamming
density of  $\phi_{j} = 0.844$. 
Thus, the lower density clogged states 
effectively phase separate into locally jammed regions 
of the point J density and regions of zero density.  
The effect of pinning on jamming can be more clearly seen 
in Fig.~6(b,c).
In Fig.~6(b) we plot the critical force $F_{c}$ 
necessary for the disks to be
depinned versus $\phi/\phi_{j}$ in the low pinning density regime
for $N_{p}/N_{j} = 0.00138$, $0.00692$,
0.0346, and 0.09267. 
At $N_{p}/N_{j} = 0.00138$ there is only a single pinning site in the
system, so $F_{c} = 0$ for $\phi/\phi_{j} < 1.0$ and $F_c$ is finite
only for $\phi/\phi_{j} > 1.0$.
In this regime, once the system depins it moves as a rigid solid.
For increasing $N_{p}/N_{j}$, the
density $\phi/\phi_{j}$ at which $F_{c}$ becomes finite decreases,
indicating that the
system is effectively jammed at lower densities. The depinning behavior is
different in character for $\phi/\phi_{j} > 1.0$ and $\phi/\phi_{j} < 1.0$. 
At the lower disk densities, depinning is associated with large plastic 
deformations, and
the $F_{c}$ vs $\phi/\phi_{j}$ curves have a peak value at the transition
from plastic to elastic depinning. 
At low pinning densities, the depinning threshold increases upon crossing
the jamming density from below; however, for higher pinning densities the
behavior is reversed.
The onset of jamming at higher pinning densities can be defined to occur when 
the system
depins rigidly or acts like an elastic solid. 
Here, the unjammed or liquidlike
state at $\phi/\phi_{j} < 1.0$ is more strongly pinned than the 
jammed or rigid state at $\phi/\phi_j>1.0$.
This is illustrated in Fig.~6(c) where we plot $F_{c}$ vs $\phi/\phi_{c}$
for $N_{p}/N_{j} = 0.828$, $0.415$, and $0.277$.
There is a sharp drop in
$F_{c}$ near $\phi/\phi_{j}$, and the system is much more strongly
pined for $\phi/\phi_{j} < 1.0$. 
A very similar phenomenon is believed to be behind 
what is called the peak effect in type-II superconducting vortex systems. 
In samples where the peak effect is observed
there are a large number of weak pinning sites. 
When the vortices form a solid structure
they are weakly pinned, while when the vortex structure is liquidlike,
the vortices can easily adjust their
positions to take advantage of the low energy pinning site positions,
and the system is strongly pinned 
\cite{S}.
This result shows that the onset of jamming can 
either increase or decrease the effectiveness of  pinning of friction depending
on the density or strength of the pinning disorder.      
Brito {\it et al.} investigated 2D jamming in systems where a 
fraction of the particles are pinned down, and found that 
as long as the packings are isostatic, 
the presence of pinning does not modify the the jamming transition \cite{54}. 

Future directions could include
examining periodic arrangements of defects, where 
the pinning length scale would be much better defined.
Additionally, for periodic substrates the jamming density may be a
function of the direction of the drift force with respect to the symmetry
directions of the substrate. 
Another approach would be to consider the effects of obstacles or pinning
when the jammed state is reached not with a drift force, but by allowing
the disk radii to expand until the energy of the relaxed system 
remains finite; properties such as the shear modulus, vibrational spectrum,
and force chains could then be examined.

\section{Systems with Longer Range Interactions}

Several works have shown that ideas from the jamming of 2D granular packings
can be applied successfully to
systems in which the particle-particle interactions are intermediate or 
long ranged, rather than short-ranged as in granular media.
For example, in charged colloidal suspensions the interactions can have
two length scales,
an intermediate range repulsive Yukawa interaction 
and a shorter range steric repulsive interaction \cite{55,56}.    
Dislocations \cite{6} and vortices in type-II superconductors \cite{57,58} 
are examples of systems with long range repulsive interactions.   
A jamming scenario for 2D gliding dislocations was recently explored 
in Ref.~\cite{6}, where the dislocations
are modeled as point particles with both attractive and 
repulsive long range strain interactions. 
Depending on the sign of its Burgers vector, 
an individual dislocation moves either in the
positive or negative x direction under an applied load.
In systems containing an equal number of dislocations with positive and
negative Burgers vectors, when a load is applied
the dislocations pile up or jam until the load exceeds a critical value,
above which the pileups can break apart.
Unlike granular media, the 
dislocations always form a jammed state at low load
even for arbitrarily low dislocation densities, and 
the critical load required to unjam
the system decreases with decreasing dislocation density but 
always remains finite \cite{6}.  
The jamming phase diagram of Nagel and Liu \cite{1} is redrawn
in Fig.~1 of Ref.~\cite{6} to show that the inverse density axis in the 
dislocation system
does not end at a point $J$ but extents out to infinity. 
Finally, the yield stress $\tau_{c}$ grows
with the dislocation density as $
\tau_c \propto \rho^{1/2}$ \cite{6}, in agreement with many experiments.  
An open question is whether the yielding transition 
for dislocations is similar to the yielding transition
for granular disk systems, or 
whether the long range interactions in the dislocation system
fundamentally change the nature of the 
yielding.   

\begin{figure}
\centering{
\includegraphics[width=3.5in]{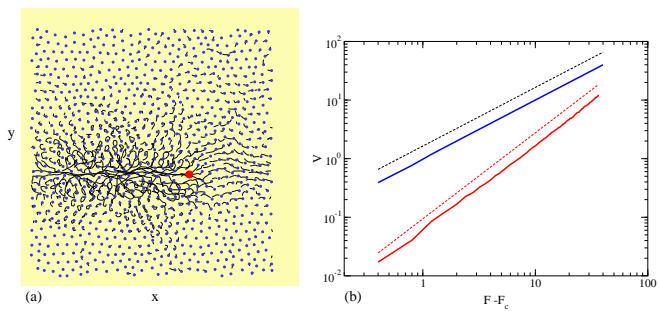}}
\caption
{(a) Image from
a 2D simulation of Yukawa interacting particles showing bidisperse background
particles (blue), the driven probe particle (red), and the
particle trajectories over a fixed time interval (black lines).
A finite threshold force is required
to move the probe particle, indicating that
the material is jammed. 
The finite threshold for motion persists 
down to very low densities due to the 
longer range interactions between the colloids. 
(b) Velocity vs force curves for the same system.
Blue solid curve: a driven particle with a small charge of $q_D=0.5$.
Blue dashed curve: a fit with $\beta=1.0$.
Red solid curve: a driven particle with a large charge of $q_D=20$.
Red dashed curve: a fit with $\beta=1.47$.
Figure adapted from Ref.~\cite{56}.
}
\label{fig:9}
\end{figure}

The fact that dislocations jam at any density is 
attributed to the long range or power law nature
of the interactions between them. 
A similar argument should apply to 
other systems with long range interactions, 
such as unscreened charged systems or
logarithmically interacting 
superconducting vortices. 
The situation can change if some type of screening 
cuts off the long range interactions. 
For instance, in polydisperse charged colloidal systems,
studies of a probe particle driven through a 3D packing found
that below the packing density at which the steric or short-range
interactions between the particles become important,
a finite threshold force must be applied to 
move the particle, 
indicating that the system is forming a weak amorphous solid \cite{55}. 
In 2D simulations of Yukawa interacting particles \cite{56} where a single
probe particle is driven through the system, there is a finite force
threshold for motion even at low densities.
In Fig.~8(a) we plot the trajectories of bidisperse Yukawa interacting
particles 
when a probe particle is driven 
though the bulk \cite{56}. 
The density is well below the jamming density of
particles with a small but finite steric interaction radius; 
nevertheless, a finite force must be applied in order to move the probe
particle, which induces strong localized motion in the surrounding
particles.
Unlike the granular jamming system 
just at point J, where rearrangements caused by a local probe particle are
spread throughout 
the entire system,
in Fig.~8(a) large rearrangements close to the driven particle serve to localize
the disturbance,
indicating that further away in the bulk the system
is acting like a solid.   
At low enough densities, the Yukawa particles are so far apart that they
no longer interact with each other, and the system is no longer jammed.
This suggests that there could be two jamming transitions
in Yukawa systems: one at low densities when the longer range 
interactions become effective, and a second transition at higher densities
that is
similar to point J for granular media, when the short range steric repulsion
interactions become important.

The question of whether systems with intermediate or long range interactions
jam can also be approached by analyzing the velocity versus force (V-F) curves 
obtained with a driven probe particle.
If the system is jammed, there should
be a finite critical depinning force $F_{c}$ required to move the
probe particle,
and the V-F curves 
may have the form
$$ V \propto (F - F_{c})^{\beta} .$$
In a viscous unjammed fluid, 
$F_{c} = 0$ and generally $\beta = 1.0$, although complex 
or glassy fluids may
have additional nonlinear 
features in the V-F curves that we do not consider here.
In Fig.~8(b) we plot the average velocity $V$ of two different sizes of 
probe particles $V$ vs $F - F_{c}$ for
the Yukawa system from Fig.~8(a). 
For the large probe particle, which induces strong plastic distortions in
the background particles when it moves,
we find $\beta = 1.47$ as indicated by the red dashed line. 
This exponent is close to that observed for plastic depinning of
vortices and colloids moving over random disorder \cite{39}. 
In a system with a lower density of particles,
$F_{c}$ decreases but the V-F scaling persists. 
For soft particles with
a finite range interaction range, once the density is low 
enough $F_{c} = 0$ and the system is unjammed. This
density corresponds to $\phi_{j}$ in the bidisperse disk system. 
If the particle-particle interactions are isotropic, it is possible to
define an effective finite particle radius $R$, and the
system could be considered jammed 
when the effective area density of
the particles reaches $0.84$. 
We note that even when the density is lower than the value at which
$F_c=0$, 
the 
V-F curves
could still 
scale as $V \propto F^\beta$; 
however, the scaling range would diminish in width
until eventually $\beta = 0$. 
For systems with infinite range interactions, 
such as dislocation assemblies, $F_{c}$ would
decrease but still be
finite for arbitrarily low 
dislocation densities. 
Another limit occurs when the charge of the probe particle
becomes so small compared to the charge of the bulk particles that
the probe particle induces virtually no plastic rearrangements in the
background particles, which act as fixed obstacles to the probe particle
motion.
In this case, $F_c$ can still be finite but the scaling
exponent becomes $\beta = 1.0$, as shown by the blue curve in 
Fig.~8(b). 
It is likely that in this limit, the system has the same behavior
as a single particle being driven over a random substrate 
where $\beta = 0.5$ very close to $F_c$ \cite{40}, with a crossover
to $\beta=1.0$ at higher drives. 
The $\beta=0.5$ behavior appears only extremely close to $F_c$, and the
measurement in Fig.~8(b) was taken outside this range.

For superconducting vortex matter, there have been numerical \cite{57} and
experimental \cite{58} studies
of 2D systems in which the vortices pass through a constricting
geometry. 
Due to the repulsive vortex-vortex interactions, the vortex motion may cease
or be strongly reduced due to jamming behavior.
This system resembles hopper flow of granular media \cite{59}, 
and demonstrates that the idea of jamming can be 
applied to other systems. 
Similar studies could be performed for the flow of 
charged colloids through funnel geometries.   

\begin{figure}
\includegraphics[width=3.5in]{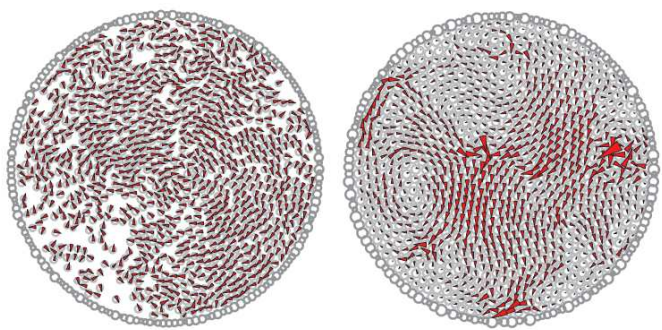}
\caption
{
Images of an active matter system in the liquid phase 
at $\phi=0.6$ (left) and in the jammed phase at $\phi=0.95$ (right), with
arrows indicating the instantaneous velocity field.
Reprinted with permission from 
S. Henkes, Y. Fily, and M.C. Marchetti, Phys.~Rev.~E {\bf 84}, 
040301(R) (2011).  Copyright 2011 by the American Physical Society.
}
\label{fig:10}
\end{figure}

\section{Active Matter and Quantum Matter}
Another class of systems that has been attracting 
much recent interest is self-propelled 
or active matter.  
Such systems could be comprised of swimming bacteria, Janus colloidal particles,
pedestrians, or traffic. 
There has already been considerable work on understanding clogging
or jamming behavior in pedestrian and traffic models \cite{3,4}. 
Here we focus on active matter made of
self-driven colloidal particles, where it has been shown that even for
systems in which the interactions between the particles are purely repulsive,
the addition of self-propulsion can produce a transition to a clustered
or self-jammed state
\cite{60,61,62,63,64}.     
Henkes {\it et al.} considered a 2D system of 
harmonically repulsive self-propelled disks 
moving according to overdamped dynamics \cite{64}, which develop 
directed motion when confined by a circular
container.
In Fig.~9 we show images from Ref.~\cite{64}
highlighting the instantaneous
particle velocities
in the low density ($\phi = 0.6$) unjammed phase and the higher density 
($\phi=0.95$) jammed phase. 
Even in the jammed phase, where the net motion of the particles
is minimal, there are still small correlated particle motions. 
This work suggests that active matter jammed phases 
might exhibit collective motions reminiscent of those found in
crawling cells.   
In a subsequent study of
self-driven polar disks with isotropic repulsive forces 
and no alignment, 
the system formed
a cluster state or phase separated state as the density or activity of the
disks increased 
\cite{64}. 
In experimental studies
of self clustering of monodisperse active particles, the clusters were termed
``living crystals''  since the
self-jammed clusters can exhibit crystalline order \cite{62}.
Although the living crystals are not strictly jammed, if such 
a system were placed in a confining environment,
the self-clustering effect could cause
the system to become effectively jammed or unable to move. 
This suggests that perhaps another axis 
for the jamming phase diagram of Ref.~\cite{1} would include a
self motility feature such as run length, 
where at very long run lengths the
system forms a cluster.

\begin{figure}
\centering{
\includegraphics[width=2.5in]{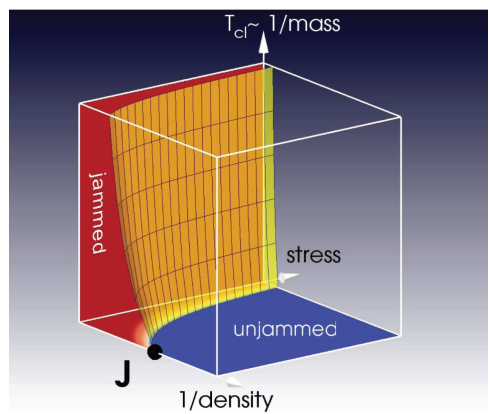}}
\caption
{Jamming phase diagram for quantum systems with axes of inverse density,
stress, and inverse mass.  
Reprinted with permission from 
Z. Nussinov, P. Johnson, M.J. Graf, and A.V. Balatsky,
Phys.~Rev.~B {\bf 87}, 
184202 (2013).  Copyright 2013 by the American Physical Society.
}
\label{fig:11}
\end{figure}

There have recently been proposals for applying jamming ideas 
to 2D or 3D quantum systems.
Nussinov {\it et al.} considered a mapping between 
classical and quantum systems to
examine quantum critical jamming of hard-core bosons 
undergoing a metal (flowing state) to insulator (jammed state) 
transition \cite{65}.
In Fig.~11 we show a schematic jamming phase diagram for quantum systems
from Ref.~\cite{65}, where at zero temperature a point J is expected
to exist.
Boninsegini {\it et al.} also argued that jamming could occur in a 
quantum system 
in a disordered assembly of $^4$He atoms, 
and that this could enhance the metastability of 
superfluid glass states \cite{66}. 
The extension of jamming concepts to quantum and semi-classical systems
is a new area and we expect 
that this will be a growing field.  

\section{Summary}
We have provided an overview of jamming in 2D systems 
focusing mostly on frictionless particle assemblies that 
form amorphous packings.
There has been considerable evidence mounting 
that at least in frictionless bidisperse disk systems, 
jamming has the properties of a 
continuous phase transition, including critical fluctuations.
Despite extensive work on this system, the exact nature of the transition
remains elusive, which could be due to a number of subtle issues. 
There is also evidence that other types of transitions, detectable
via dynamic response,
could be occurring below the jamming point J, 
while for high densities above point J a new type of deep
jamming behavior can occur. 
We discuss jamming in 2D systems with varied particle shapes, such as
an assembly of chains which jams at significantly lower densities than the
disk systems.
New types of jamming or clogging behaviors can occur 
for disk assemblies in the presence
of obstacles or pinning sites, suggesting that there may 
be a new axis, quenched disorder, on the jamming phase diagram. 
The ideas of jamming 
in 2D can also be applied to systems with 
longer range interactions such as dislocations, 
vortices in superconductors, and charged colloids,
and it remains an open question whether jamming in these systems 
has the same universal properties as jamming in disk systems.
Active matter or self-driven particle systems is another area in which
ideas of jamming can be applied, and recent work suggests that yet
another axis, activity, can be described and can produce dynamic jamming in
an active matter jamming
phase diagram.
Ideas of jamming are also beginning to be applied to quantum systems. 
Jamming has become a field in its own right 
as the ideas of jamming can be applied to a wide variety of 
situations, and jamming should continue
to be an active and growing field for the foreseeable future.

\section{Acknowledgments}
This work was carried out under the auspices of the 
NNSA of the 
U.S. DoE
at 
LANL
under Contract No.
DE-AC52-06NA25396.


\footnotesize{

\end{document}
\begin{thebibliography}{99}

\bibitem{1}
A. J. Liu and S. R. Nagel, {\it Nature (London)}, 1998, {\bf 396}, 21;
A.J. Liu and S.R. Nagel, {\it Ann. Rev. Condens. Matter Phys.}, 2010, {\bf 1},
347.


\bibitem{2} 
B. S. Kerner and H. Rehborn, 
{\it Phys. Rev. Lett.}, 1997, {\bf 79}, 4030.

\bibitem{3}
D. Helbing,
{\it Rev. Mod. Phys.}, 2001,  {\bf 73}, 1067.

\bibitem{4}
D. Helbing, I.J. Farkas, and T. Vicsek, {\it Nature (London)}, 2000, {\bf 407},
487.

\bibitem{5}
C. Leduc, K. Padberg-Gehle, V. Varge, D. Helbing, S. Diaz, and J. Howard,
{\it Proc. Natl. Acad. Sci. (USA)}, 2012, {\bf 109}, 6100.

\bibitem{6}
G. Tsekenis, N. Goldenfeld, and K.A. Dahmen,
{\it Phys. Rev. Lett.}, 2011, {\bf 106}, 105501.

\bibitem{7}
K. Harada, O. Kamimura, H. Kasai, T. Matsuda, A. Tonomura, 
and V. V. Moshchalkov, {\it Science}, 1996, {\bf 274}, 1167.


\bibitem{8}
C.S. O’Hern, L.E. Silbert, A.J. Liu, and S.R. Nagel, 
{\it Phys. Rev. E}, 2003, {\bf 68}, 011306.

\bibitem{9}
M.E. Cates, J.P. Wittmer, J.-P. Bouchaud, and P. Claudin,
{\it Phys. Rev. Lett.}, 1998, {\bf 81}, 1841.

\bibitem{10}
M. Boninsegni, L. Pollet, N. Prokof’ev, and B. Svistunov,
{\it Phys. Rev. Lett.}, 2012, {\bf 109}, 025302;
Z. Nussinov, P. Johnson, M.J. Graf, and A.V. Balatsky,
{\it Phys. Rev. B}, 2013, {\bf 87}, 184202;
D. Poletti, P. Barmettler, A. Georges, and C. Kollath,
{\it Phys. Rev. Lett.}, 2013, {\bf 111}, 195301.

\bibitem{11}
N. Goldenfeld,
{\it Lectures on Phase Transitions and the Renormalization Group}
(Westview Press, Oxford, 1992).

\bibitem{12}
D. Vagberg, D. Valdez-Balderas, M.A. Moore, P. Olsson, and S. Teitel,
{\it Phys. Rev. E}, 2011, {\bf 83}, 030303(R).

\bibitem{13}
H. Hinrichsen, {\it Adv. Phys.}, 2000, {\bf 49}, 815;
G. Odor, {\it Rev. Mod. Phys.}, 2004, {\bf 76}, 663.

\bibitem{14.1}
K.A. Takeuchi, M. Kuroda, H. Chate, and M. Sano, 
{\it Phys. Rev. Lett}, 2007, {\bf 99}, 234503.

\bibitem{28}
L. Corte, P.M. Chaikin, J.P. Gollub, and D.J. Pine, 
{\it Nature Phys.}, 2008, {\bf 4}, 420.

\bibitem{14}
A. Donev, S. Torquato, F.H. Stillinger, and R. Connelly,
{\it Phys. Rev. E}, 2004, {\bf 70}, 043301.

\bibitem{15}
C.F. Schreck, C.S. O'Hern, and L.E. Silbert,
{\it Phys. Rev. E}, 2011, {\bf 84}, 011305.

\bibitem{16}
L.E. Silbert, A.J. Liu, and S.R. Nagel, {\it Phys. Rev. Lett.},
2005, {\bf 95}, 098301.

\bibitem{16.N}
M. Dennin, {\it J. Phys.: Condens. Matter}, 2008, {\bf 20}, 283103.

\bibitem{17}
J.A. Drocco, M.B. Hastings, C.J. Olson Reichhardt, and C. Reichhardt,
{\it Phys. Rev. Lett.}, 2005, {\bf 95}, 088001.

\bibitem{18}
E. Kolb, P. Cixous, N. Gaudouen, and T. Darnige,
{\it Phys. Rev. E}, 2013, {\bf 87}, 032207.

\bibitem{19}
P. Olsson and S. Teitel, {\it Phys. Rev. Lett.}, 2007, {\bf 99}, 178001.


\bibitem{20}
C. Heussinger and J.-L. Barrat,
{\it Phys. Rev. Lett.}, 2009, {\bf 102}, 218303.
	
\bibitem{21}
D.A. Head, {\it Phys. Rev. Lett.}, 2009, {\bf 102}, 138001.

\bibitem{22}
M. Wyart, S.R. Nagel, and T.A. Witten, {\it Europhys. Lett.},
2005, {\bf 72}, 486.

\bibitem{23}
J.M. Schwarz, A.J. Liu, and L.Q. Chayes, {\it Europhys. Lett.},
2006, {\bf 73}, 560.

\bibitem{24}
S. Henkes and B. Chakraborty, {\it Phys. Rev. Lett.}, 2005, {\bf 95}, 
198002.

\bibitem{25}
C.P. Goodrich, W.G. Ellenbroek, and A.J. Liu,
{\it Soft Matter}, 2013, {\bf 9}, 10993;
S.S. Schoenholz, C.P. Goodrich, O. Kogan, A.J. Liu and S.R. Nagel,
{\it Soft Matter}, 2013, {\bf 9}, 11000.

\bibitem{wait}
S.R. Waitukaitis, L.K. Roth, V. Vitelli, and H.M. Jaeger,
{\it EPL}, 2013, {\bf 102}, 44001.

\bibitem{24.N} 
W.G. Ellenbroek, E. Somfai, M. van Hecke, and W. van Saarloos, 
{\it Phys. Rev. Lett.}, 2006, {\bf 97}, 258001.

\bibitem{24.N1}
M. Mailman and B. Chakraborty,
{\it J. Stat. Mech.: Theory Exp.}, 2011, {\bf 2011}, L07002.

\bibitem{ohern2002}
C. O'Hern, S.A. Langer, A.J. Liu, and S.R. Nagel,
{\it Phys. Rev. Lett.}, 2002, {\bf 88}, 075507.

\bibitem{katgert2010}
G. Katgert and M. van Hecke,
{\it EPL}, 2010, {\bf 92}, 34002.

\bibitem{26}
T.S. Majmudar, M. Sperl, S. Luding, and R.P. Behringer, 
{\it Phys. Rev. Lett.}, 2007, {\bf 98}, 058001.

\bibitem{newhall2013}
J. Newhall, J. Cao, and S.T. Milner,
{\it Phys. Rev. E}, 2013, {\bf 87}, 052203.

\bibitem{27}
E.J. Banigan, M.K. Illich, D.J. Stace-Naughton, D.A. Egolf,
{\it Nature Physics}, 2013, {\bf 9}, 288.

\bibitem{keys2007}
A.S. Keys, A.R. Abate, S.C. Glotzer, and D.J. Durian,
{\it Nature Phys.}, 2007, {\bf 3}, 260.

\bibitem{heu2010}
C. Heussinger, P. Chaudhuri, and J.-L. Barrat,
{\it Soft Matter}, 2010, {\bf 6}, 3050.

\bibitem{29}
C.F. Schreck, R.S. Hoy, M.D. Shattuck, and C.S. O’Hern,
{\it Phys. Rev. E}, 2013, {\bf 88}, 052205.

\bibitem{30}
T. Shen, C.S. O'Hern, and M.D. Shattuck,
{\it Phys. Rev. E}, 2012, {\bf 85}, 011308.

\bibitem{31}
P. Olsson and S. Teitel
{\it Phys. Rev. E}, 2013, {\bf 88}, 010301.

\bibitem{32}
C. Zhao, K. Tian, and N. Xu,
{\it Phys. Rev. Lett.}, 2011, {\bf 106}, 125503.

\bibitem{33}
K.W. Desmond, P.J. Young, D. Chen, and E.R. Weeks,
{\it Soft Matter}, 2013, {\bf 9}, 3424.

\bibitem{34}
G. Katgert, B.P. Tighe, and M. van Hecke,
{\it Soft Matter}, 2013, {\bf 9}, 9739.

\bibitem{35}
D. Bi, J. Zhang, B. Chakraborty, and R.P. Behringer,
{\it Nature}, 2011, {\bf 480}, 355.

\bibitem{36}
L.E. Silbert,
{\it Soft Matter}, 2010, {\bf 6}, 2918.

\bibitem{37}
J. Geng and R.P. Behringer, {\it Phys. Rev. E}, 2005, {\bf 71}, 011302.

\bibitem{38}
C.J. Olson Reichhardt and C. Reichhardt,
{\it Phys. Rev. E}, 2010, {\bf 82}, 051306.

\bibitem{39}
D. Dom{\' i}nguez, {\it Phys. Rev. Lett.}, 1994, {\bf 72}, 3096;
C. Reichhardt, C.J. Olson, N. Gr{\ o}nbech-Jensen, and F. Nori,
{\it Phys. Rev. Lett.}, 2001, {\bf 86}, 4354;
C. Reichhardt and C.J. Olson,
{\it Phys. Rev. Lett.}, 2002, {\bf 89}, 078301;
Y. Fily, E. Olive, N. Di Scala, and J.C. Soret,
{\it Phys. Rev. B}, 2010, {\bf 82}, 134519;
S.-Z. Lin, C. Reichhardt, C.D. Batista, and A. Saxena,
{\it Phys. Rev. B}, 2013, {\bf 87}, 214419.

\bibitem{40}
D.S. Fisher, {\it Phys. Rev. B}, 1985, {\bf 31}, 1396.

\bibitem{41}
R. Candelier and O. Dauchot, {\it Phys. Rev. E}, 2010, {\bf 81}, 011304.

\bibitem{42}
A. Fiege, M. Grob, and A. Zippelius,
{\it Granular Matter}, 2012, {\bf 14}, 247.

\bibitem{43}
T. Boerzsoenyi and R. Stannarius,
{\it Soft Matter}, 2013, {\bf 9}, 7401.

\bibitem{44}
E. Ben-Naim, Z.A. Daya, P. Vorobieff, and R.E. Ecke,
{\it Phys. Rev. Lett.}, 2001, {\bf 86}, 1414.

\bibitem{prentis2002}
J.J. Prentis and D.R. Sisan,
{\it Phys. Rev. E}, 2002, {\bf 65}, 031306.

\bibitem{46}
E. Brown, A. Nasto, A.G. Athanassiadis, and H.M. Jaeger,
{\it Phys. Rev. Lett.}, 2012, {\bf 108}, 108302.

\bibitem{45}
L.-N. Zou, X. Cheng, M.L. Rivers, H.M. Jaeger, and S.R. Nagel, 
{\it Science}, 2009, {\bf 326}, 408.

\bibitem{47}
C.J. Olson Reichhardt and L.M. Lopatina, {\it Science}, 2009, {\bf 326}, 374.

\bibitem{48}
L.M. Lopatina, C.J. Olson Reichhardt, and C. Reichhardt,
{\it Phys. Rev. E}, 2011, {\bf 84}, 011303.

\bibitem{zhang2010}
J. Zhang, T.S. Majumdar, A. Tordesillas, and R.P. Behringer,
{\it Granular Matter}, 2010, {\bf 12}, 159.

\bibitem{49}
C.J. Olson, C. Reichhardt, M. McCloskey, and R.J. Zieve, 
{\it EPL}, 2002, {\bf 57}, 904.

\bibitem{50}
S.J. Gerbode, D.C. Ong, C.M. Liddell, and I. Cohen,
{\it Phys. Rev. E}, 2010, {\bf 82}, 041404.

\bibitem{51}
C.F. Schreck, N. Xu, and C.S. O'Hern,
{\it Soft Matter}, 2010, {\bf 6}, 2960.

\bibitem{52}
M. Mailman, C.F. Schreck, C.S. O'Hern, and B. Chakraborty,
{\it Phys. Rev. Lett.}, 2009, {\bf 102}, 255501.

\bibitem{53}
C.J. Olson Reichhardt, E. Groopman, Z. Nussinov, and C. Reichhardt,
{\it Phys. Rev. E}, 2012, {\bf 86}, 061301.

\bibitem{S}
S. Bhattacharya and M.J. Higgins,
{\it Phys. Rev. Lett.}, 1993, {\bf 70}, 2617.
	
\bibitem{54}
C. Brito, G. Parisi, and F. Zamponi,
{\it Soft Matter}, 2013, {\bf 9}, 8540.

\bibitem{55}
P. Habdas, D. Schaar, A.C. Levitt, and E.R. Weeks,
{\it Europhys. Lett.}, 2004, {\bf 67}, 477.

\bibitem{56}
M.B. Hastings, C.J. Olson Reichhardt, and C. Reichhardt,
{\it Phys. Rev. Lett.}, 2003, {\bf 90}, 098302.

\bibitem{57}
C.J. Olson Reichhardt and C. Reichhardt,
{\it Phys. Rev. B}, 2010, {\bf 81}, 224516.

\bibitem{58}
G. Karapetrov, V. Yefremenko, G. Mihajlovic, J. E. Pearson, 
M. Iavarone, V. Novosad, and S.D. Bader,
{\it Phys. Rev. B}, 2012, {\bf 86}, 054524.

\bibitem{59}
C.C. Thomas and D.J. Durian,
{\it Phys. Rev. E}, 2013, {\bf 87}, 052201.

\bibitem{60}
Y. Fily and M. C. Marchetti, {\it Phys. Rev. Lett.}, 2012, {\bf 108}, 
235702.

\bibitem{61}
G.S. Redner, M.F. Hagan, and A. Baskaran,
{\it Phys. Rev. Lett.}, 2013, {\bf 110}, 055701.

\bibitem{62}
J. Palacci, S. Sacanna, A.P. Steinberg, D.J. Pine, and P.M. Chaikin,
{\it Science}, 2013, {\bf 339}, 936.

\bibitem{63}
I. Buttinoni, J. Bialke, F. K{\" u}mmel, H. L{\" o}wen, 
C. Bechinger, and T. Speck,
{\it Phys. Rev. Lett.}, 2013, {\bf 110}, 238301.

\bibitem{64}
S. Henkes, Y. Fily, and M.C. Marchetti,
{\it Phys. Rev. E}, 2011, {\bf 84}, 040301.

\bibitem{65}
Z. Nussinov, P. Johnson, M.J. Graf, and A.V. Balatsky,
{\it Phys. Rev. B}, 2013, {\bf 87}, 184202.

\bibitem{66}
M. Boninsegni, L. Pollet, N. Prokof’ev, and B. Svistunov,
{\it Phys. Rev. Lett.}, 2012, {\bf 109}, 025302.

\end{thebibliography}
